\documentclass[12pt, draftclsnofoot, onecolumn]{IEEEtran}
\IEEEoverridecommandlockouts
\usepackage{cite}
\usepackage{amsmath,amssymb,amsfonts}
\usepackage[linesnumbered,ruled,vlined]{algorithm2e}
\usepackage{graphicx}
\usepackage{epstopdf}
\usepackage{textcomp}
\usepackage{xcolor}
\usepackage{multirow}
\usepackage{soul}
\usepackage{array}

\usepackage[justification=centering]{caption}   

\usepackage{booktabs}

\usepackage{booktabs}
\def\BibTeX{{\rm B\kern-.05em{\sc i\kern-.025em b}\kern-.08em
    T\kern-.1667em\lower.7ex\hbox{E}\kern-.125emX}}
\begin{document}

\title{High Stable and Accurate Vehicle Selection Scheme Based on Federated Edge Learning in Vehicular Networks}



\author{

{ 
	Qiong Wu,~\IEEEmembership{Senior Member,~IEEE}, Xiaobo Wang, Qiang Fan,\\ Pingyi Fan, ~\IEEEmembership{Senior Member,~IEEE}, Cui Zhang, Zhengquan Li
}

\thanks{

{	

	Qiong Wu and Xiaobo Wang are with the School of Internet of Things Engineering, Jiangnan University, Wuxi 214122, China, and also with the State Key Laboratory of Integrated Services Networks (Xidian University),  Xi'an 710071, China (e-mail: qiongwu@jiangnan.edu.cn, xiaobowang@stu.jiangnan.edu.cn)

	Qiang Fan is with Qualcomm, San Jose, CA 95110, USA (e-mail: qf9898@gmail.com)

	Pingyi Fan is with the Department of Electronic Engineering, Beijing National Research Center for Information Science and Technology, Tsinghua University, Beijing 100084, China (Email: fpy@tsinghua.edu.cn)

	Cui Zhang is with Banma Network Technology Co., Ltd., Shanghai 200000, China

	Zhengquan Li is with the School of Internet of Things Engineering, Jiangnan University, Wuxi 214122, China, and also with Jiangsu Future Networks Innovation Institute, Nanjing 211111, China
}

}
}

\maketitle

\begin{abstract}
Federated edge learning (FEEL) technology for vehicular networks is considered as a promising technology to reduce the computation workload while keeping the privacy of users. In the FEEL system, vehicles upload data to the edge servers, which train the vehicles' data to update local models and then return the result to vehicles to avoid sharing the original data. However, the cache queue in the edge is limited and the channel between edge server and each vehicle is time-varying. Thus, it is challenging to select a suitable number of vehicles to ensure that the uploaded data can keep a stable cache queue in edge server while maximizing the learning accuracy. Moreover, selecting vehicles with different resource statuses to update data will affect the total amount of data involved in training, which further affects the model accuracy. In this paper, we propose a vehicle selection scheme, which maximizes the learning accuracy while ensuring the stability of the cache queue, where the statuses of all the vehicles in the coverage of edge server are taken into account. The performance of this scheme is evaluated through simulation experiments, which indicates that our proposed scheme can perform better than the known benchmark scheme.
\end{abstract}


\section{introduction}
\label{Introduction}
With the development of the internet of vehicles (IoV) and big data technologies \cite{1, 2, 3, 4, 5}, different vehicles can share the data collected by on-board sensors to a central cloud, which then trains the model based on the collected data \cite{6}, thus meeting the requirement of different vehicle applications, such as autonomous driving, location awareness, object/gesture recognition, mobile biometrics and mobile augmented reality \cite{7, 8, 9, 10, 11, 12}. However, vehicle data are often private and different vehicles are reluctant to share their own data with other vehicles, thus the central cloud is hard to achieve an accurate model due to the shortage of training data \cite{55, 56, 57, 58, 59, 60, 61}. To solve this problem, federated learning is introduced to protect vehicle privacy by sharing models rather than sharing the raw data of vehicles \cite{13, 14}. Specifically, each vehicle gets a local model by training the data locally and then uploads the local model to the central cloud. The central cloud aggregates the uploaded local models to get the global model and then sends the global model to vehicles \cite{15, 16}. In this way, the vehicle can get an accurate model while protecting its data. However, local training in federated learning can impose a large computational burden on vehicles with limited computational power \cite{17, 18, 19}. 

\begin{figure}[htbp]
    \centering
    \includegraphics[width=1\textwidth]{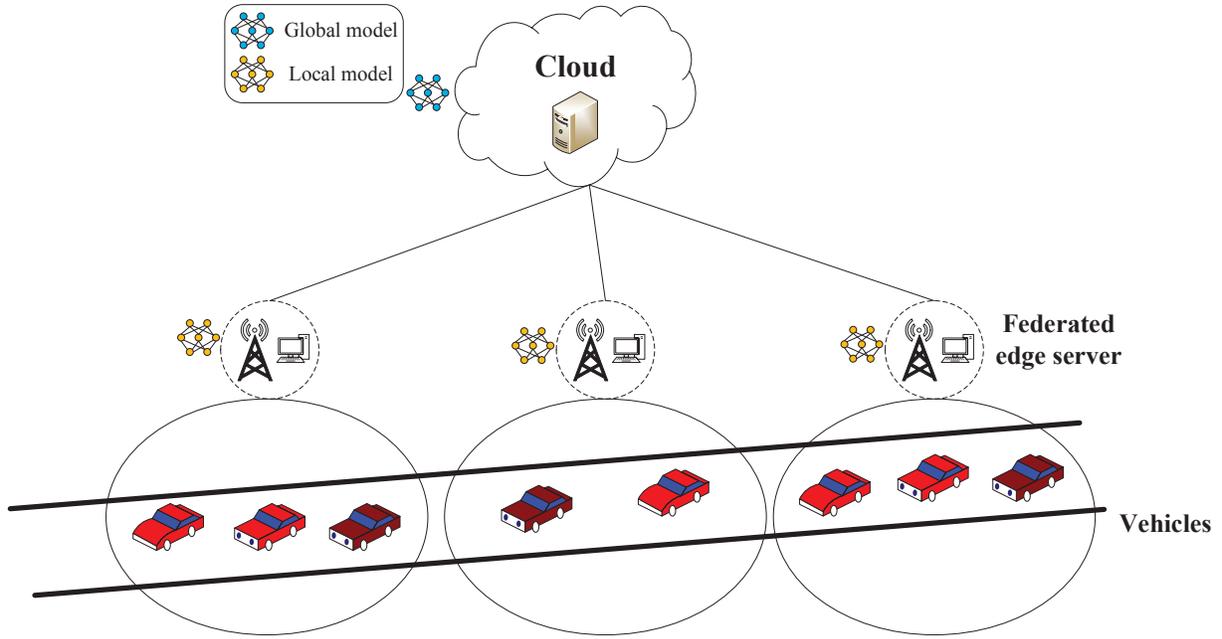}
    \caption{The FEEL system.}
	\label{fig1}
\end{figure}

In this paper, we consider a three-tier infrastructure including cloud, edge servers and vehicles. In edge servers, federated edge learning (FEEL) is employed. It can relieve vehicles' burden and protect vehicle data between the vehicle and the central cloud. Specifically, the edge server collects and trains the vehicles' data to get a local model, and then returns the result to vehicles to avoid sharing the original data. Afterwards, the central cloud collects and aggregates all local models from the edge servers to obtain the global model, then distributes the global model to the edge servers and the vehicles for their use, thus protecting the privacy of the vehicles while avoiding the computation workload on the vehicle \cite{20, 21}. The FEEL system is shown in Figure~\ref{fig1}. Note that in the FEEL system, the mobile edge computing (MEC) servers would not harm the users' privacy through deploying a shuffler between the vehicles and each edge server, which can disrupt the data uploaded by the vehicles.

In the FEEL system, the data uploaded by vehicles for training are temporarily stored in the cache queue of each edge server. The cache queue of each edge server is limited. For each edge server, if too many vehicles upload data to it, the cache queue becomes overflowed and unstable. If too few vehicles upload data to it, there will be insufficient data in the cache queue for training and thus deteriorating the accuracy of models. To get accurate models while keeping the queue stable, vehicles' data should try to fill up the cache queue of the edge server. On the other hand, the training starts right after the local data arrives at the cache queue. Once the training is finished, the training results are sent back to the selected vehicles and the used training data will be deleted from the cache queue only when the channel condition is good enough. In this case, it indicates that data randomly leaves the cache queue accordingly. It is challenging to select a suitable number of vehicles to upload the data to the cache queue where the used data and the result leave randomly.
In addition, vehicles in the network have different resource statuses including remaining data amount, communication quality, remaining energy and survivability (i.e., time duration within the coverage of an edge server). Thus, considering the resource statuses to select vehicles is critical. To the best of our knowledge, there is no related work to improve the accuracy of the model by considering different vehicles' resource statuses and queue stability for vehicle selection in federated learning, which motivated us to carry out this work.

In this paper, we propose a vehicle selection scheme in federated learning, with edge server which can control the cache queue stability by determining the number of vehicles selected for uploading and selects vehicles to maximize the learning accuracy while ensuring the stability of the cache queue in it\footnote {Code to replicate the numerical experiments here presented can be found at https://github.com/qiongwu86/-Vehicle-selection.}. The main contributions of this paper are shown as follows
\begin{itemize}
\item[1)]We consider a FEEL system in vehicular networks and propose a vehicle selection scheme to maximize the accuracy of the training models while ensuring the stability of the cache queue.

\item[2)]For each edge server, we formulate the problem based on the Lyapunov theorem to estimate the number of vehicles which can get the high accuracy of models while keeping the queue stable.

\item[3)]We consider four resource statuses of vehicles including remaining data amount, communication quality, remaining energy and survivability, and propose an algorithm to select vehicles to improve the accuracy of models.
\end{itemize}

The rest of this paper is organized as follows. Section~\ref{RELATED WORK} reviews the related work. Section~\ref{SYSTEM MODEL} briefly describes the system model. Section~\ref{PROBLEM FORMULATION} formulates the problem of determining the optimal number of selected vehicles. Section~\ref{VEHICLE SELECTION SCHEME} proposes a vehicle selection scheme to select vehicles. We give simulation results in Section~\ref{SIMULATION RESULTS AND ANALYSIS} and conclude the paper in Section~\ref{CONCLUSION}.

\section{RELATED WORK}
\label{RELATED WORK}
In this section, we review the existing research work on the edge-assisted federated learning system. We first review the work related to the edge-assisted federated learning system where users train models locally, then review the work on the FEEL system where the edge servers assist users in model training.

In \cite{22}, Ren et al. proposed an edge-assisted federated learning system where each user is equipped with a Central Processing Unit (CPU) or Graphics Processing Unit (GPU) to speed up the training process, which overcomes serious privacy issues and long communication latency due to the heavy traffic towards the centralized node. In \cite{23}, Mohammad et al. regarded wireless access nodes as parameter servers (PS) in the edge-assisted federated learning systems, where users update their local model based on their local data and the global model which is received from the PS. After the local updates, users transfer their updated model to the PS over a wireless fading multiple access channel (MAC), which enables reliable transmission from users to PS. In \cite{24}, Zhu et al. designed a low-latency multi-access scheme for the edge-assisted federated learning systems to address communication latency issues and accelerate the learning process. It is proposed that the updates simultaneously transmitted by devices over broadband channels should be aggregated over the air by exploiting the waveform superposition property of a multi-access channel. In \cite{25}, Liu et al. studied over-the-air model aggregation in the edge-assisted federated learning system, where channel state information at the transmitters (CSIT) is unavailable. It leverages the reconfigurable intelligent surface (RIS) technology to align the cascaded channel coefficients for CSIT-free model aggregation. In \cite{26}, Taïk et al. considered data characteristics in wireless scheduling algorithms in the edge-assisted federated learning system, which allocates bandwidth based on user data to minimize the training time and total transmission energy while maximizing user diversity. In \cite{27}, Mo et al. exploit the combination of communication and computation in the edge-assisted federated learning, in which both the communication resource allocation for global machine learning (ML) parameters aggregation and the computation resource allocation for locally updating ML parameters are jointly optimized, thus minimizing the total energy consumption of the user in a limited training time. In \cite{28}, Zhu et al. proposed an efficient broadband analog transmission scheme in the federated learning system to overcome a communication bottleneck due to the resource sharing of edge devices. In \cite{29}, Ren et al. studied gradient averaging over participating devices in each round of communication in the federated learning system. They also proposed a scheduling policy to exploit both diversity in multiuser channels and the different weights of model updates from different edge devices, thus achieving the optimal trade-off between channel quality and model weights. In \cite{18}, Zeng et al. proposed power devices using wireless power transfer (WPT) in order to enable some energy-constrained devices to execute some power-hungry learning tasks. In \cite{30}, Wang et al. studied an edge-based communication optimization framework in the federated learning system to reduce the number of the end devices which is directly connected to the parameter server thereby avoiding high latency and unnecessary communication. In \cite{31}, Lin et al. studied a social federated edge learning (SFEL) framework over wireless networks, where the potential social attributes among edge devices and their users can be exploited. They built a social graph model to recruit the trustworthy social friends as learning partners by considering the mutual trust and learning task similarity, thus enhancing the learning performance as well as the stability of the edge-assisted federated learning system. In \cite{32}, Albaseer et al. studied a federated learning model where the edge server is equipped with multiple antennas. They focused on searching for the optimal user resources, including the fine-grained selection of relevant training samples, bandwidth, transmission power, beamforming weights and processing speed to minimize the total energy consumption given a deadline constraint on the communication rounds. In \cite{33}, He et al. developed an importance-aware joint data selection and resource allocation algorithm in the edge-assisted federated learning system, which enables the user to select important data to improve learning efficiency. In \cite{34}, Su et al. proposed a dynamic data and channel adaptive sensor scheduling power control algorithm based on a residual feedback mechanism. The derived decentralized optimal solution can be used to compute both the channel state information and data importance to seize good transmission opportunities and important gradients. In \cite{35}, Sun et al. proposed an energy-aware dynamic device scheduling algorithm to optimize the training performance by considering an over-the-air edge-assisted federated learning system with analog gradient aggregation. In \cite{36}, Liu et al. focused on reducing the learning latency due to the limited local computing power and communication bandwidth. They introduced model pruning for wireless federated learning systems to reduce the neural network scale and maximize the convergence rate under the given learning latency budget via jointly optimizing the pruning ratio, device selection, and wireless resource allocation. In \cite{37}, Zhou et al. studied how to optimize the transmission power and bandwidth allocation based on user quality of experience (QoE). They proposed a low energy consumption bandwidth allocation algorithm, which allocates appropriate power to the device based on the computing power and channel state of the device, thereby reducing energy consumption. In \cite{38}, Cao et al. employed the transmission power control to reduce aggregation errors in the edge-assisted federated learning system. Different from conventional power control designs, they considered a new power control design aiming at directly maximizing the convergence speed, thus minimizing the mean square error. In \cite{39}, Fan et al. studied the over-the-air model aggregation in an edge-assisted federated learning system. They developed a message-passing-based algorithm to fulfill this estimation task with low complexity and near-optimal performance. In \cite{40}, Xiao et al. studied to select the appropriate vehicles to participate in learning tasks and optimize resource allocation under learning time and energy consumption constraints. By taking the vehicle position and velocity into consideration, they formulated a min-max optimization problem to jointly optimize the onboard computation capability, transmission power and local model accuracy to achieve the minimum cost in the worst case of FL. In \cite{41}, Nishio et al. studied a mobile edge computing (MEC) framework for machine learning (ML) technologies. They proposed a new FL protocol to mitigate the problem that the overall training process becomes inefficient for the client due to the limited computational resources or poor wireless channel conditions. In \cite{42}, Bao et al. proposed an edge computing-based joint client selection and networking scheme for vehicular IoT. The scheme assigns some vehicles as edge vehicles by employing a distributed approach, and uses the edge vehicles as FL clients to conduct the local model training, thus the optimal behaviors can be learned based on the interaction with environments. However, the above works only enable users to train local models, which will increase the communication overhead and computational burden of users.

In \cite{15}, Luo et al. studied a three-tier FEEL system containing user, edge and cloud layers, where the edge server helps the user to do model training. They proposed an efficient resource scheduling algorithm to solve the problem of joint allocation of computing and communication resources for users, to minimize the consumed resources. In \cite{43}, Mhaisen et al. leveraged the edge computing paradigm to design a hierarchical FEEL system to solve the problem of poor global models due to the over-fitting of local models to local data. The system executes federated gradient descent on the user-edge layer and federated averaging on the edge-cloud layer. In this hierarchical architecture, they formalized and optimized this user-edge assignment problem such that edge-level data distributions turn to be similar, which enhances the federated averaging performance. However, although the users of the above-related works carry out model training through the help of servers to alleviate the communication overhead and the computational burden of users, none of them considered the selection of users in vehicular networks, which motivated us to start this work. 

\begin{figure*}
\center
\includegraphics[width=1\textwidth]{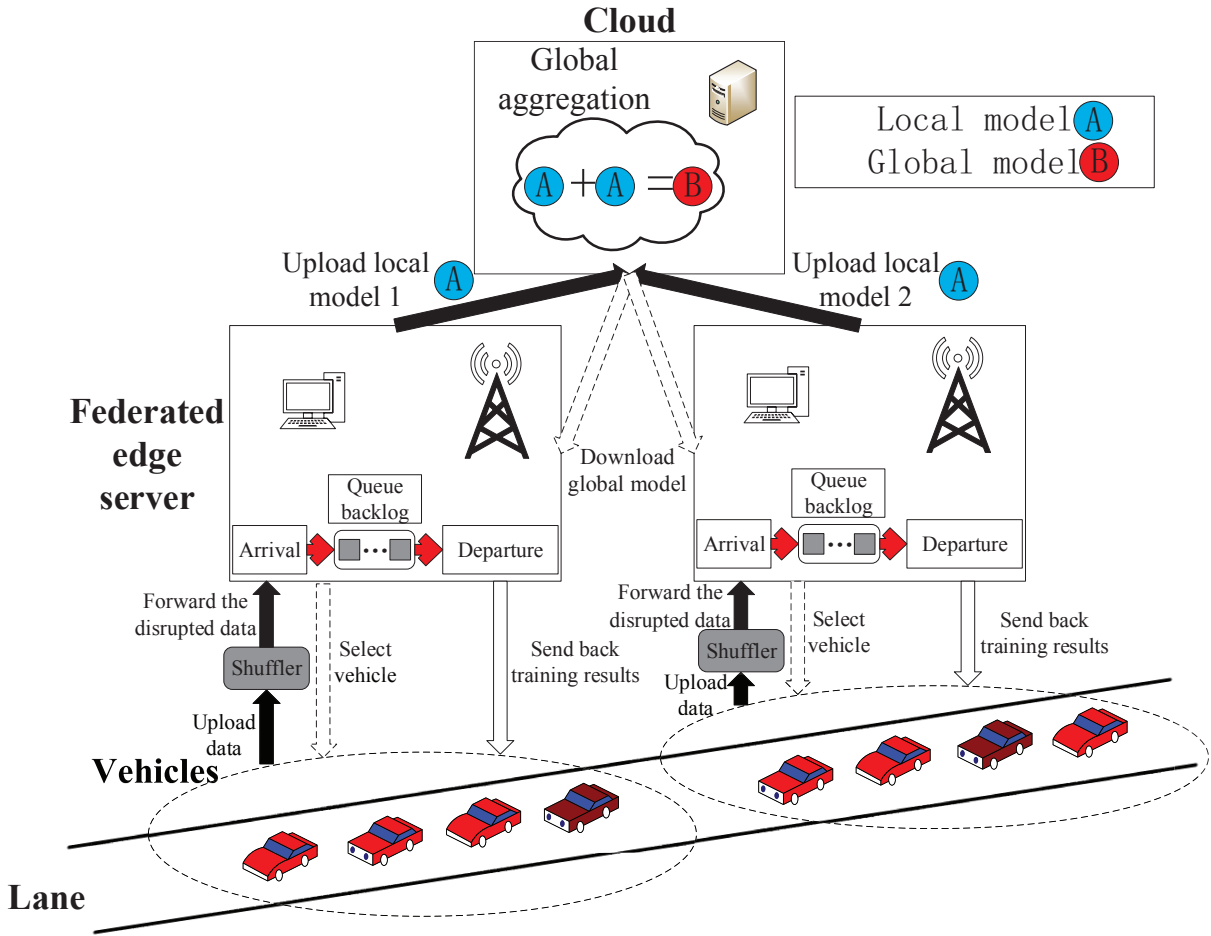}
\caption{System model figure.}
\label{fig2}
\end{figure*}

\section{SYSTEM MODEL}
\label{SYSTEM MODEL}
In this paper, we consider a FEEL system as shown in Figure~\ref{fig2}. The system consists of vehicles, edge servers, and a central cloud. The edge servers are deployed along the road. The coverage of each server is a circular area with a radius of $R$. Each edge server is equipped with a cache to store the data waiting for training, where the cache adopts the first-come, first-served (FCFS) policy with a maximum queue length of $Q_{max}$. If the data stored in the cache queue exceeds the queue limit, the data will be discarded due to overflow. It is assumed that $K$ vehicles are traveling in the same direction with a uniform speed in the coverage area, and the speeds of different vehicles follow a truncated Gaussian distribution, which is flexible and consistent with the real dynamic vehicular environment. Note that in the FEEL system, the edge servers would not harm the vehicles' privacy through deploying a shuffler between the vehicles and each edge server, which can disrupt the data uploaded by the vehicles. The duration time that vehicles drive in the coverage of each edge server is divided into equal time slots. In each time slot, each vehicle in the coverage of the edge server calculates the resource statuses including remaining data amount, communication quality, remaining energy and survivability, then transmits its resource statuses to the edge server. Afterwards, the edge server will select some vehicles to upload data for training. Specifically, the edge server first estimates the number of selected vehicles based on the backlog of the current cache queue to satisfy the queue without overflowing and then determines which vehicles are selected according to the vehicles' resource statuses to ensure that more data are involved in the training. After that, the selected vehicles first upload data to the shuffler, then the shuffler disrupts all the received data in a uniformly random permutation, afterwards the shuffler forwards the disrupted data to the edge server. The training starts right after data are collected, i.e., the data arrive at the cache queue of the edge server. The edge server updates the local models of the selected vehicles when the training is finished. Then, if the communication channel conditions between some selected vehicles and the edge server are good enough, the edge server sends back the training results to the corresponding selected vehicles and deletes the used training data from the cache queue. Otherwise, the used training data will be backlogged in the cache queue until the communication channel conditions between the edge server and some selected vehicles become good enough. Then the next time slot is started and the above process is repeated in order to update local model. When the vehicles in the coverage area of the edge server have no data to upload, the edge server uses these periods of time to upload the updated local model to the central cloud. When all edge servers upload their local models, the central cloud can aggregate the local models to achieve a global model and then feed back the new update global model to each edge server. Table~\ref{tab1} lists the notations in this paper.

\section{PROBLEM FORMULATION}
\label{PROBLEM FORMULATION}
In this section, we will adopt the Lyapunov control theorem to formulate the problem of determining the optimal number of selected vehicles based on the backlog of the cache queue.

\begin{table*}[htbp]
	\caption{Notations\label{tab:table1}}
	\centering
	\label{tab1}
	\centering
	\setlength{\tabcolsep}{2mm}{
	\begin{tabular}{|c|c|c|c|}
		\hline
		\textbf{Notation} & \textbf{Description} & \textbf{Notation} & \textbf{Description}  \\
		\hline
		$K$ & The total number of vehicles. & $Noise^k(t)$ &\multicolumn{1}{m{6.6cm}|}{The value to measure the noise over the channel for vehicle $k$ in time slot $t$.} \\
		\hline
		$t_0$ & The initial time slot.  & $N$ &The path loss exponent. \\
		\hline
		$D$ &The coverage range of an edge server. & $f$ &The carrier frequency. \\
		\hline
		$R$ &The radius of edge server coverage area. & $f_{d}^k(t)$ &The Doppler frequency of vehicle $k$ in time slot $t$. \\
		\hline
		$Q_{max}$ & The maximum length of the cache queue. & ${dist}^k(t)$ &\multicolumn{1}{m{6.6cm}|}{The distance from the vehicle $k$ to the edge server in time slot $t$.} \\
		\hline
		$Q(t)$ &The backlog of the cache queue size in time slot $t$. & $m^k(t)$ &\multicolumn{1}{m{6.6cm}|}{The radio wave propagation distance of vehicle $k$ in time slot $t$.} \\
		\hline
		$U(n(t))$ &\multicolumn{1}{m{6.6cm}|}{The utility function of expected accuracy when the number of selected vehicles is $n(t)$ in time slot $t$.} & $v^k$ &The speed of vehicle $k$. \\
		\hline
		$\lambda(n(t))$ &\multicolumn{1}{m{6.6cm}|}{The amount of data arriving the edge server when the number of vehicles is $n(t)$ in time slot $t$.}  & $B$ &The wavelength. \\
		\hline
		$\mu(t)$ &\multicolumn{1}{m{6.6cm}|}{The amount of data leaving the edge server in time slot $t$.}  & $\theta^k(t)$ &\multicolumn{1}{m{6.6cm}|}{The angle between the driving direction of vehicle $k$ in time slot $t$ and the direction of radio wave incidence.}  \\
		\hline
		$\boldsymbol{X}$ &The set of vehicles numbers. & $h$ &The edge server height. \\
		\hline
		$V$ &\multicolumn{1}{m{6.6cm}|}{The trade-off factor between accuracy and queue-backlog.}  & $r^k(t)$ &The transmission rate of vehicle $k$ in time slot $t$. \\
		\hline
		$n(t)^*$ &\multicolumn{1}{m{6.6cm}|}{The optimal number of selected vehicles in time slot $t$.}  & $\alpha$ &The vehicle connection status. \\
		\hline
		$n(t)$ &The number of selected vehicles in time slot $t$. & $\delta^2$ &The variance of the estimated signal strength. \\
		\hline
		$C_d^k(t)$ &\multicolumn{1}{m{6.6cm}|}{The remaining data amount of vehicle $k$ in time slot $t$.}  & $\varepsilon_k(t)$ &\multicolumn{1}{m{6.6cm}|}{The correlation between vehicle $k$ and the edge server.} \\
		\hline
		$C_{com}^k(t)$ &\multicolumn{1}{m{6.6cm}|}{The communication quality of vehicle $k$ in time slot $t$.}  & $E_{tran}^k(t)$ &\multicolumn{1}{m{6.6cm}|}{The transmission energy consumption of vehicle $k$ in time slot $t$.} \\
		\hline
		$C_E^k(t)$ & The remaining energy of vehicle $k$ in time slot $t$. & $C_n$ &\multicolumn{1}{m{6.6cm}|}{The data size transmitted in each time slot.} \\
		\hline
		$C_S^k(t)$ &The survivability of vehicle $k$ in time slot $t$.   & $C_{sur}^k(t_0)$ &\multicolumn{1}{m{6.6cm}|}{The survivability of vehicle $k$ in the initial time slot $t_0$.} \\
		\hline
		$P^k(t)$ &The transmitting power of vehicle $k$ in time slot $t$.  & $d^k$ &\multicolumn{1}{m{6.6cm}|}{The initial position of vehicle $k$ in the coverage area of the edge server.} \\
		\hline
		$L^k(t)$ &\multicolumn{1}{m{6.6cm}|}{The corresponding path loss of power for vehicle $k$ in time slot $t$.}  & $w^k(t)$ &The priority weight of vehicle $k$ in time slot $t$. \\
		\hline
		
	\end{tabular}}
\label{tab1}
\end{table*}

\begin{figure}[htbp]
    \centering
    \includegraphics[width=1\textwidth]{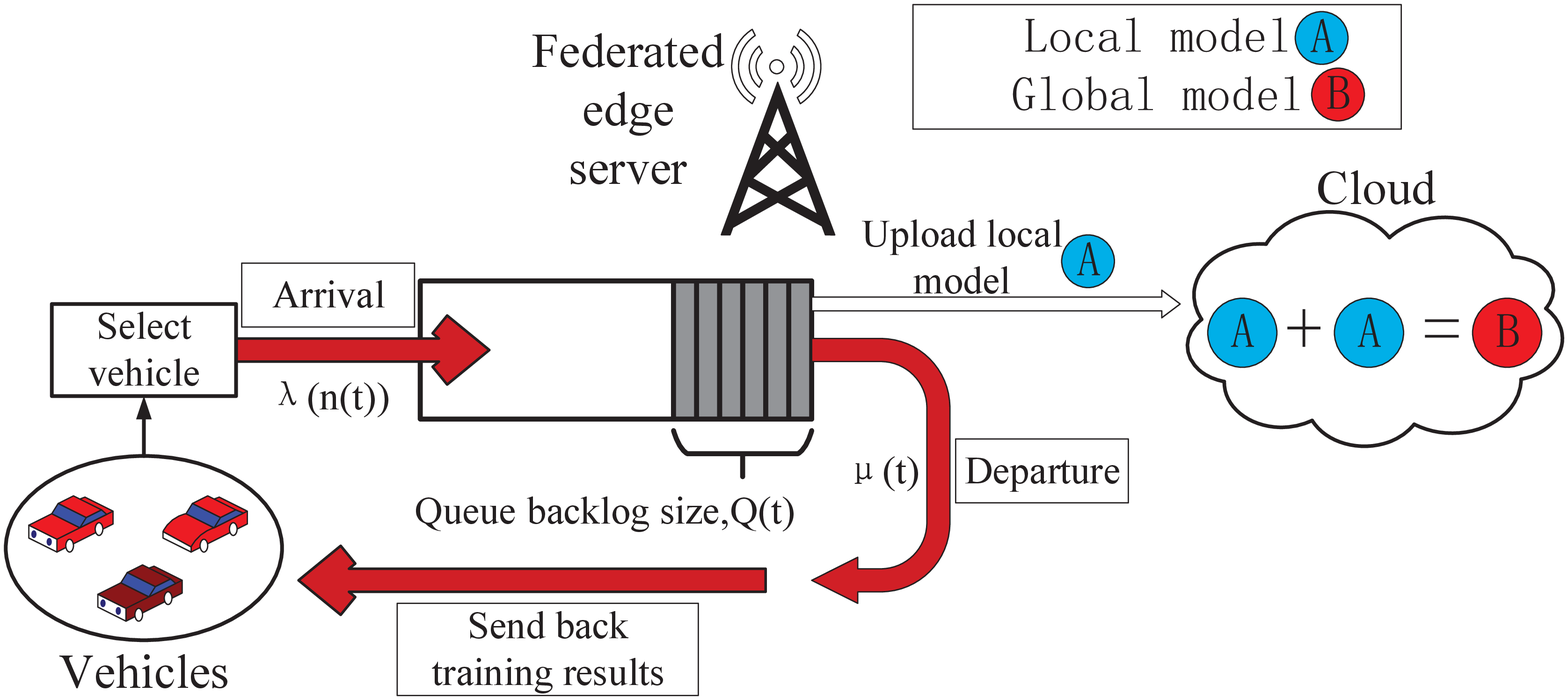}
    \caption{Edge server cache queue model figure.}
	\label{fig3}
\end{figure}

From queue theory, the backlog of the cache queue in one time slot is determined by the arrival and departure of the previous time slot, as shown in Figure~\ref{fig3}. The cache queue of the edge server can be represented as
\begin{equation}
Q(t+1) = \max \{Q(t)+\lambda(t)-\mu(t),0\},
\label{eq1}
\end{equation}
where $Q(t)$ is the backlog of the cache queue in time slot $t$, $\lambda(t)$ is the amount of data arriving at the edge server in time slot $t$ and $\mu(t)$ is the amount of data leaving the edge server in time slot $t$.

For each edge server target is to maximize the learning accuracy, i.e., select as many as possible vehicles to upload data, under the constraint that the backlog size of the cache queue of each edge server does not exceed the maximum queue length. According to the Lyapunov optimization theorem \cite{44}, this problem can be formulated as
\begin{equation}
P_1: \max : \lim\limits_{T\rightarrow \infty}\frac{1}{T}\sum_{t=0}^{T-1}(U(n(t)),
\label{eq2}
\end{equation}
\begin{equation}
s.t.\quad \lim\limits_{T\rightarrow \infty}\frac{1}{T}\sum_{t=0}^{T-1}Q(t)\leq Q_{max},
\label{eq3}
\end{equation}
where $U(n(t))$ is the utility function of expected accuracy when the number of selected vehicles is $n(t)$ in time slot $t$, $T$ is the number of time slots and $Q_{max}$ is the maximum queue length. Eq.~\eqref{eq3} is the queue stability constraint.

Since $P_1$ is a long-term optimization problem within multiple time slots, solving $P_1$ to obtain the optimal number of vehicles is non-trivial. The short-term control decisions between adjacent time slots affect significantly on the long-term problem, therefore we transform $P_1$ based on the Lyapunov theorem to obtain a short-term optimization problem and thus the problem can be solved simply.

The Lyapunov equation is a scalar, non-negative representation of the backlog of a queue within the current time slot $t$ \cite{44}, which is defined as
\begin{equation}
L(\Theta(t))\triangleq\frac{1}{2}Q(t)^2,
\label{eq4}
\end{equation}
where $\Theta(t)$ is defined as a vector to measure $Q(t)$, i.e., $\Theta(t)\triangleq Q(t)$.

In Lyapunov theorem, Lyapunov drift is adopted to optimize the queue stability \cite{44}. Lyapunov drift is defined as the growth volume of a queue backlog between adjacent time slots, i.e., 
\begin{equation}
\bigtriangleup(\Theta(t)) \triangleq L(\Theta(t+1))-L(\Theta(t)).
\label{eq5}
\end{equation}

In this paper, we consider optimizing the queue stability while optimizing the utility function of expected accuracy, thus we add the utility function within the current slot, i.e., $U(n(t))$, to both sides of Eq.~\eqref{eq5} to obtain the optimization objective, i.e., drift plus-penalty term, which is written as
\begin{equation}
P_2: \max : \bigtriangleup(\Theta(t))+V\cdot U(n(t)),
\label{eq6}
\end{equation}
\begin{equation}
s.t.\quad \lim\limits_{T\rightarrow \infty}\frac{1}{T}\sum_{t=0}^{T-1}Q(t)\leq Q_{max},
\label{eq7}
\end{equation}
where $V$ is a non-negative trade-off parameter to make the tradeoff between the utility function and the Lyapunov drift.

In this paper, $\bigtriangleup(\Theta(t))$ is affected by $\lambda(t)$ and $\mu(t)$, next we will further derive the relationships between $\bigtriangleup(\Theta(t))$ and $\lambda(t)$, $\mu(t)$ to relax this optimization problem.

We first derive $Q(t)^2$ to determine $L(\Theta(t))$ in Eq.~\eqref{eq4}. According to Eq.~\eqref{eq1}, if $Q(t)+\lambda(t)-\mu(t)\geq 0$, then $Q(t+1)=Q(t)+\lambda(t)-\mu(t)$, thus we have
\begin{equation}
Q(t+1)^2 = (Q(t)+\lambda(t)-\mu(t))^2.
\label{eq8}
\end{equation}

Otherwise, if $Q(t)+\lambda(t)-\mu(t)\textless 0$, then $Q(t+1)=0\textgreater Q(t)+\lambda(t)-\mu(t)$, so
\begin{equation}
Q(t+1)^2 = 0\textless (Q(t)+\lambda(t)-\mu(t))^2.
\label{eq9}
\end{equation}

Combining Eq.~\eqref{eq8} and Eq.~\eqref{eq9}, we have
\begin{equation}
Q(t+1)^2 \leq (Q(t)+\lambda(t)-\mu(t))^2.
\label{eq10}
\end{equation}

Substituting Eq.~\eqref{eq10} into Eq.~\eqref{eq4}, we have
\begin{equation}
\begin{aligned}
&L(\Theta(t+1))=\frac{1}{2}Q(t+1)^2 \\ &\leq\frac{1}{2}Q(t)^2+\frac{1}{2}(\lambda(t)-\mu(t))^2+Q(t)(\lambda(t)-\mu(t)).
\label{eq11}
\end{aligned}
\end{equation}

Substituting Eq.~\eqref{eq4} and Eq.~\eqref{eq11} in to  Eq.~\eqref{eq5}, we have
\begin{equation}
\begin{aligned}
\bigtriangleup(\Theta(t)) &= \frac{1}{2}Q(t+1)^2-\frac{1}{2}Q(t)^2\\ &\leq\frac{1}{2}(\lambda(t)-\mu(t))^2+Q(t)(\lambda(t)-\mu(t))\\ &\leq C+Q(t)(\lambda(t)-\mu(t)),
\end{aligned}
\label{eq12}
\end{equation}
where the $\frac{1}{2}(\lambda(t)-\mu(t))^2$ term is positive and can not reflect the varying direction of $\lambda(t)-\mu(t)$, it can be upper bounded by a constant $C$ \cite{44}.

Substituting Eq.~\eqref{eq12} in to  Eq.~\eqref{eq6}  we have
\begin{equation}
\begin{aligned}
&\bigtriangleup(\Theta(t))+V\cdot U(n(t))\\ &\leq C+V\cdot U(n(t))+Q(t)(\lambda(t)-\mu(t)).
\end{aligned}
\label{eq13}
\end{equation}

From Eq.~\eqref{eq13}, to relax the optimization problem $P_2$ in Eq.~\eqref{eq6}, one can consider the following optimization problem:
\begin{equation}
P_3: \max : V\cdot U(n(t))+Q(t)(\lambda(t)-\mu(t)),
\label{eq14}
\end{equation}
\begin{equation}
s.t.\quad \lim\limits_{T\rightarrow \infty}\frac{1}{T}\sum_{t=0}^{T-1}Q(t)\leq Q_{max}.
\label{eq15}
\end{equation}

We consider each selected vehicle transmits the same data size to the edge server, thus the amount of data arriving the edge server is related to the number of selected vehicles $n(t)$. Let $\lambda(n(t))$ be the amount of data arriving the edge server when the number of selected vehicles is $n(t)$, which can be calculated as
\begin{equation}
\lambda(n(t))=C_n \times n(t),
\label{eq16}
\end{equation}
where $C_n$ is the transmitted data size for each selected vehicle in each time slot. In this case, the optimization objective of problem $P_3$ can be expressed as
\begin{equation}
\begin{aligned}
&n^*(t)\leftarrow\\ &\mathop{\arg\max}\limits_{n(t)\in \boldsymbol{X}}\{V\cdot U(n(t))+Q(t)\cdot\left[\lambda(n(t))-\mu(t)\right]\},
\end{aligned}
\label{eq17}
\end{equation}
where $n^*(t)$ is the optimal number of selected vehicles in time slot $t$,  $\boldsymbol{X}$ is the set of the numbers of the vehicles that can be selected ($\boldsymbol{X}=\{0,1,2,\cdots, K\}$). Since the edge server returns the training result to the vehicles only when the communication channel state is good enough and the wireless channel condition is random in each time slot, thus $\mu(t)$ is a random value.

According to the problem formulated above, the deterministic optimal number of selected vehicles $n^*(t)$ can be calculated by the Lyapunov control theorem according to Eqs.~\eqref{eq1}-\eqref{eq17}, hence the target to estimate the optimal number of selected vehicles is formulated as finding the optimal $n^*(t)$ to maximize the optimization objective Eq.~\eqref{eq17} under the constraint Eq.~\eqref{eq15}.

\section{VEHICLE SELECTION SCHEME}
\label{VEHICLE SELECTION SCHEME}

In this section, we propose a vehicle selection scheme to select vehicles. Specifically, in each time slot $t$, each vehicle in the coverage of the edge server calculates the resource statuses including remaining data amount, communication quality, remaining energy and survivability, and then transmits its resource statuses to the edge server. Afterwards, the edge server selects vehicles according to its backlog of the cache queue and the resource statuses of vehicles. Next, we will first introduce how vehicles calculate the resource statuses, then introduce the proposed algorithm for selecting vehicles.

\subsection{Resource statuses}
Note that each time slot is long enough to ensure each vehicle could calculate the resource statuses. The resource statuses include remaining data amount, communication quality, remaining energy and survivability.

\subsubsection{Remaining data amount}

This type of resource indicates the amount of data carried by the vehicle $k$ in time slot $t$, which is denoted as $C^k_d(t)$. It assumed that each vehicle could know its remaining data amount in real-time.

\subsubsection{Communication quality} 

According to  \cite{20}, the communication quality of each vehicle $k$ in each slot $t$ is mainly affected by the receiver receiving power and noise, which can be calculated as
\begin{equation}
C_{com}^k(t) = \left[P^k(t) - L^k(t)\right]\times Noise^k(t),
\label{eq18}
\end{equation}
where $P^k(t)$ is the transmitting power of vehicle $k$ in time slot $t$ in $dB$, $Noise^k(t)$ is a value to measure the noise over the channel for vehicle $k$ in time slot $t$ and $L^k(t)$ is the corresponding power of path loss for vehicle $k$ in time slot $t$. 

In this paper, we consider the international tecommunication union-radiocommunication sector (ITU-R) path loss model, thus $L^k(t)$ in Eq.~\eqref{eq18} is calculated as \cite{45}
\begin{equation}
L^k(t) = 20\lg \left[f+f_d^k(t)\right]+10N\lg\left[dist^k(t)\right]-28,
\label{eq19}
\end{equation}
where $f$ is the carrier frequency, $dist^k(t)$ is the distance from vehicle $k$ to the edge server in time slot $t$, $N$ is the path loss exponent, $f_d^k(t)$ is the Doppler frequency of vehicle $k$ in time slot $t$ \cite{46}, which can be calculated as

\begin{equation}
f_d^k(t) = \frac{v^k}{B}\cos\theta^k(t),
\label{eq20}
\end{equation}
where $v^k$ is the speed of vehicle $k$, $\theta^k(t)$ is the angle between the driving direction of vehicle $k$ in time slot $t$ and the direction of radio wave incidence, and $B$ is the wavelength, which can be calculated respectively as
\begin{equation}
B=\frac{c}{f},
\label{eq21}
\end{equation}
\begin{equation}
\cos \theta^k(t)=\frac{dist^k(t)}{m^k(t)},
\label{eq22}
\end{equation}
where $c$ is the speed of light and $m^k(t)$ is the radio wave propagation distance of vehicle $k$ in time slot $t$ can be expressed as
\begin{equation}
m^k(t)=\sqrt{(h)^2+(dist^k(t))^2},
\label{eq23}
\end{equation}
where $h$ is the edge server height.

In Eq.~\eqref{eq20}, $v^k$ is generated by a truncated Gaussian distribution with the probability density function shown as follows \cite{47}:

\begin{equation}
f({v^k}) = \left\{ \begin{aligned}
\frac{{{e^{ - \frac{1}{{2{\sigma ^2}}}{{({v^k} - \mu )}^2}}}}}{{\sqrt {2\pi {\sigma ^2}} (erf(\frac{{{v_{\max }} - \mu }}{{\sigma \sqrt 2 }}) - erf(\frac{{{v_{\min }} - \mu }}{{\sigma \sqrt 2 }}))}},\\
{v_{min }} \le {v^k} \le {v_{max }},\\
0 \qquad \qquad \qquad \qquad \quad otherwise,
\end{aligned} \right.
\label{eq24}
\end{equation}
where $\sigma^{2}$ is the variance of $v^k$, $\mu$ is the mean of $v^k$, $erf\left(\frac{v^k-\mu}{\sigma \sqrt{2}}\right)$ is the Gauss error function of $v^k$ under the variance $\sigma^{2}$ and mean $\mu$, $v_{\min}$ and $v_{\max}$ are the minimum and maximum speed limit of each vehicle, respectively, thus $v^k$ can be determined under the given $\sigma^{2}$, $\mu$, $v_{\min}$ and $v_{\max}$.

Moreover, we consider that vehicles adopt wireless Transmission Control Protocol/Internet Protocol (TCP/IP) to transmit data and the transmission keeps steady state in each slot, thus $P^k(t)$ in Eq.~\eqref{eq18} can be calculated as \cite{48}
\begin{equation}
P^k(t)=\left[\frac{r^k(t)}{\alpha}\right]^2,
\label{eq25}
\end{equation}
where $\alpha$ is a value to reflect the state of vehicular connection and $r^k(t)$ is the transmission rate of vehicle $k$ in time slot $t$, which can be calculated as \cite{49}
\begin{equation}
r^k(t) = \delta^2a^k(t),
\label{eq26}
\end{equation}
where $\delta^2$ is the variance of the estimated signal strength, which is a value in the range of $[3,10] dB^2$, $a^k(t)$ is the correlation coefficient of shadow fading model for vehicle $k$ in time slot $t$, which can be calculated as \cite{49}
\begin{equation}
a^k(t) = \varepsilon_k(t)^{v^k/dist^k(t)},
\label{eq27}
\end{equation}
where $\varepsilon_k(t)$ is the correlation between vehicle $k$ and the edge server, which can be obtained according to the shadow fading model in \cite{49}.

Thus, given $c$, $h$, $\alpha$, $\delta^2$, $f$, $N$ and $v^k$, each vehicle $k$ can calculate the communication quality in time slot $t$, i.e., $C_{com}^k(t)$, according to Eqs.~\eqref{eq18}-\eqref{eq27}.

\subsubsection{Remaining energy}

Since the power consumption that vehicles download the training model from the edge server is small enough, it is negligible as compared to the power consumption that vehicles transmit data in the uplink \cite{40}. In this case, the remaining energy of vehicle $k$ in time slot $t$, denoted as $C_E^k(t)$ can be calculated as the difference between the remaining energy of vehicle $k$ in time slot $t-1$ and the energy consumption caused by transmission in time slot $t$, denoted as $E_{tran}^k(t)$, thus $C_E^k(t)$ can be expressed as
\begin{equation}
C_E^k(t) = C_E^k(t-1) - E_{tran}^k(t).
\label{eq28}
\end{equation}

We consider each vehicle transmits the same amount of data to edge server in each slot, thus the energy consumption caused by transmission for vehicle $k$ in time slot $t$ is calculated as \cite{40}
\begin{equation}
E_{tran}^k(t) = \frac{C_n\sqrt{P^k(t)}}{\alpha},
\label{eq29}
\end{equation}
where $P^k(t)$ can be calculated according to Eqs.~\eqref{eq25}- ~\eqref{eq27} under the given parameters $\alpha$, $\delta^2$ and $v^k$.

Thus, given $C_n$, $\alpha$ and the energy of vehicle $k$ in the initial time slot $t_0$, i.e., $C_E^k(t_0)$, each vehicle $k$ can calculate the remaining energy in each time slot $t$ according to Eqs.~\eqref{eq28} and ~\eqref{eq29}.

\subsubsection{Survivability}

The survivability of vehicle $k$ in time slot $t$, denoted as $C_S^k(t)$, is the time duration  within the coverage of an edge server for vehicle $k$ in time slot $t$, which is reduced by a unit time after each time slot, thus we have
\begin{equation}
C_S^k(t) = C_S^k(t-1) - t.
\label{eq30}
\end{equation}

In the initial time slot $t_0$, the survivability of vehicle $k$ can be calculated as
\begin{equation}
C_{sur}^k(t_0) = (D-d^k)/v^k,
\label{eq31}
\end{equation}
where $D$ is the coverage range of an edge server, $d^k$ is the initial position of vehicle $k$.

Thus, given $D$, $d^k$ and $v^k$, each vehicle $k$ can calculate the survivability in each time slot $t$ according to Eqs.~\eqref{eq30} and ~\eqref{eq31}.

According to the calculations of the resource statuses, the deterministic values of the resource statuses can be calculated according to Eqs.~\eqref{eq18}-\eqref{eq31}. After each vehicle calculates its resource statuses, it will transmit the resource statuses to the edge server for vehicle selection.

\begin{algorithm}
  \caption{Solution algorithm}
  \KwIn{$\boldsymbol {C}_n$}
  \KwOut{$\boldsymbol{A(t)}$}
  \label{al1}
  \textbf{initialize} {$\boldsymbol{N}_t = \boldsymbol{W}_t = \varnothing$}\\
  {$Q(t) \gets 0$}\;
  {$n^*(t) \gets 0$}\;
  {$n(t) \gets 0$}\;
  \For{\textup{each time slot} $t$}
  {
    \textup{The edge server receives $C_d^k(t)$, $C_{com}^k(t)$, $C_{E}^{k}(t)$ and $C_{S}^{k}(t)$ from each vehicle $k$}\;
    \For{\textup{each} n(t) = \textup{0,1,...,} K $(n(t) \in \boldsymbol{X})$}
    {
     \If{$0\leq Q(t-1)+\lambda(n(t)) \leq Q_{max}$}
     {$Q^* \gets - \infty$\;
    	{$Q \gets V \cdot U(n(t)) + Q(t) \cdot \left[\lambda(n(t)) - \mu(t)\right]$}\;
	{$Q^* \gets Q$\;
    }
    }
    \ElseIf{$Q(t-1)+\lambda(n(t)) \textgreater Q_{max}$}
    {$n^*(t) \gets n(t)-1$\;
    \textbf{break}}
    }

  \For{\textup{each vehicle} k = \textup{1, 2, ...,} K}
  {
       \If{$C_S^k(t)\quad $\bf{and}$ \quad C_E^k(t)\neq 0$}
       {$w^{k}(t)=\frac{C_{d}^{k}(t) \cdot C_{com}^{k}(t)}{C_{E}^{k}(t) \cdot C_{S}^{k}(t)}$}
       \Else{$w^k(t) \gets 0$}
	  {$\boldsymbol{W}_t = \boldsymbol{W}_t \cup w^{k}(t)$}
   }
   Sort $\boldsymbol{W}_t$ in descending order\;
   \For{\textup{each} $i=1,2, ..., n^*(t) (W_t[i] \in \boldsymbol{W}_t)$} 
   {
   	$\boldsymbol{N}_t$ = $\boldsymbol{N}_t \cup W_t[i]$
   }
   \For{\textup{each vehicle} $k \in \boldsymbol{N}_t$} 
   {
   	vehicle upload data $C_n$
   }
   \textbf{Training data}\;
   output training accuracy $A(t)$\;
   {$Q(t+1) \leftarrow \max \{Q(t)+\lambda(t)-\mu(t), 0\}$}
  }
\end{algorithm}

In this subsection, we will propose an algorithm to select vehicles in each time slot. Specifically, the edge server first estimates the number of selected vehicles based on the backlog of the current buffer queue to satisfy the queue without overflowing and then estimates which vehicles are selected based on the priority of each vehicle, which is calculated according to the vehicle's resource statuses, to ensure that a larger amount of data are involved in training. The input of the algorithm is the transmitted data size for each vehicle $C_n$. The output of the algorithm is the training accuracy $A$. The pseudocode of the algorithm is illustrated in Algorithm~\ref{al1}. Next, we will describe the algorithm in detail.

Let $\boldsymbol{N}_t$ be the set of the selected vehicles' number in time slot $t$, $\boldsymbol{W}_t$ be the set of the priorities of all vehicles in time slot $t$. We set the initial parameters at the beginning of the algorithm. Specifically, $\boldsymbol{N}_t$ and $\boldsymbol{W}_t$ are set to be empty, $Q(t)$, $n^*(t)$ and $n(t)$ are set to 0 (lines 1-4). Next, vehicles will be selected in each time slot $t$. 

In each time slot $t$, the edge server first receives $C_d^k(t)$, $C_{com}^k(t)$, $C_{E}^{k}(t)$ and $C_{S}^{k}(t)$ from each vehicle $k$, then it estimates the optimal number of selected vehicles through iteration. As mentioned in Section~\ref{PROBLEM FORMULATION} the target is finding the optimal $n^*(t)$ to maximize the optimization objective Eq.~\eqref{eq17} under the constraint Eq.~\eqref{eq15}. For each iteration, if the backlog of the queue after data arrive at the queue does not exceed the maximum length of queue size, i.e., $0\leq Q(t-1)+\lambda(n(t)) \leq Q_{max}$, the maximum value of the optimization objective $Q^*$ is set as the negative infinity, then the value of the optimization objective $Q$ is calculated according to Eq.~\eqref{eq17}, then $Q^*$ is set as $Q$ and $n(t)$ is increased by one to start the next iteration. Otherwise, if $Q(t-1)+\lambda(n(t))>Q_{max}$, $n^*(t)$ is set as $n(t)-1$ and the iteration is terminated. Thus, $n^*(t)$ can be determined(lines 6-16).

Then the edge server determines which vehicles should be selected. Specifically, the edge server first calculates the priority of each vehicle $k$ based on its resource statuses. The priority is defined according to the events in the system, which are described as follows. In the system, if vehicles with a small remaining data amount are selected, some vehicles with a large remaining data amount may have no chance to update all amounts of data. Moreover, vehicles would transmit a large amount of data when the communication quality is good. In addition, since the vehicles with small remaining energy and survivability would have no chance to be selected within a short time, they should be selected with high priority. Therefore, if $C^k_S(t)$ and $C^k_E(t)$ is not 0, the priority of vehicle $k$ is calculated as
\begin{equation}
w^k(t) = \frac{C^k_d(t)\cdot C^k_{com}(t)}{C^k_E(t)\cdot C^k_S(t)}.
\label{eq32}
\end{equation}

Otherwise, $w^k(t)$ is 0. Then the calculated priorities of all vehicles are stored in  $\boldsymbol{W}_t$, thus $\boldsymbol{W}_t=\{w^1(t),w^2(t),...,w^K(t)\}$ (lines 17-24). Then the edge server sorts the priorities in $\boldsymbol{W_t}$ in descending order (line 25). 

\begin{figure*}[htbp]
    \centering
    \includegraphics[width=1\textwidth]{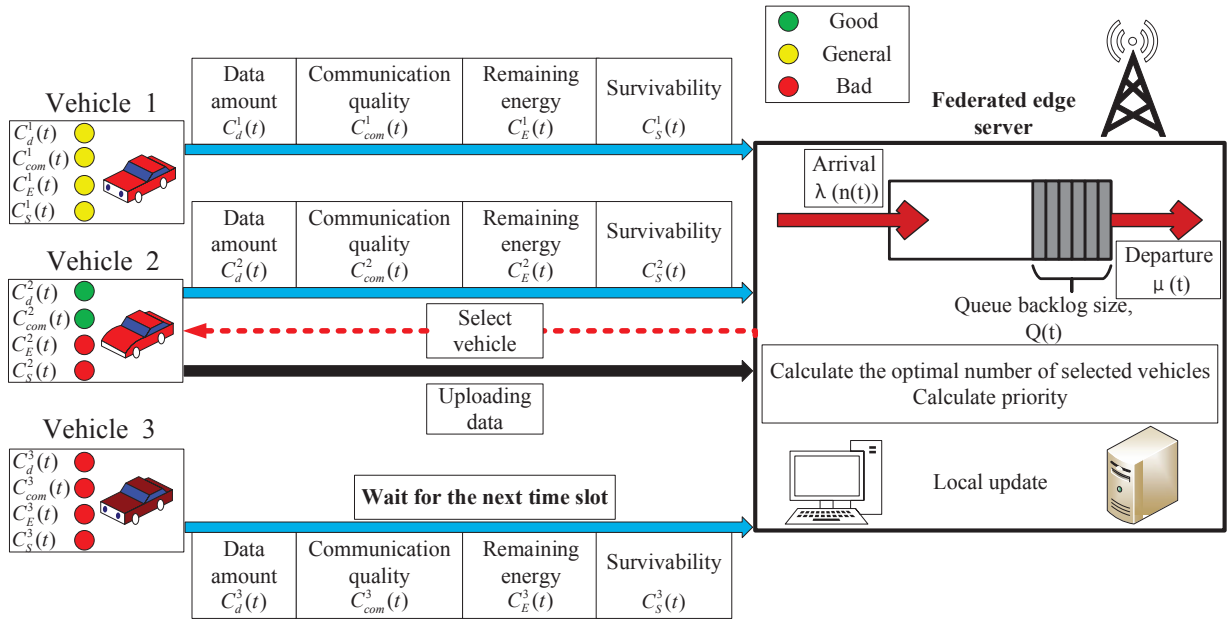}
    \caption{The process that edge server selects vehicles.}
	\label{fig4}
\end{figure*}

Afterwards, the edge server stores the vehicles' number in $\boldsymbol{N}_t$ for the vehicles with the $n^*(t)$ largest priorities in $\boldsymbol{W}_t$, which is corresponding to lines 26-28, where $W_t[i]$ is the vehicle's number corresponding to the $i$th priority in $\boldsymbol{W_t}$. Then, the edge server selects the vehicles whose numbers are stored in $\boldsymbol{N}_t$ to upload their data(lines 29-31). After receiving the data of selected vehicles, the edge server uses a two-dimensional convolutional neural network to train a local model and output the training accuracy $A(t)$(lines 32-33). Finally, the edge server updates the queue backlog according to Eq.~\eqref{eq1} and starts to select vehicles and calculate the training accuracy in the next slot time(line 34).

The process that edge server selects vehicles is illustrated in Figure~\ref{fig4} for easy understanding, where three vehicles driving in the coverage of the considered edge server. In Figure~\ref{fig4}, vehicle 2 is selected with high priority due to its good resource statuses, and vehicle 1 is selected with low priority due to its general resource statuses, while the resource statuses of vehicle 3 are bad, thus vehicle 3 is not selected and should wait for the next time slot. Note that all the values of parameters in the above selection process are deterministic, thus we can obtain a deterministic optimal set of vehicles.

\section{SIMULATION RESULTS AND ANALYSIS}
\label{SIMULATION RESULTS AND ANALYSIS}

\begin{figure}[htbp]
    \centering
    \includegraphics[width=1\textwidth]{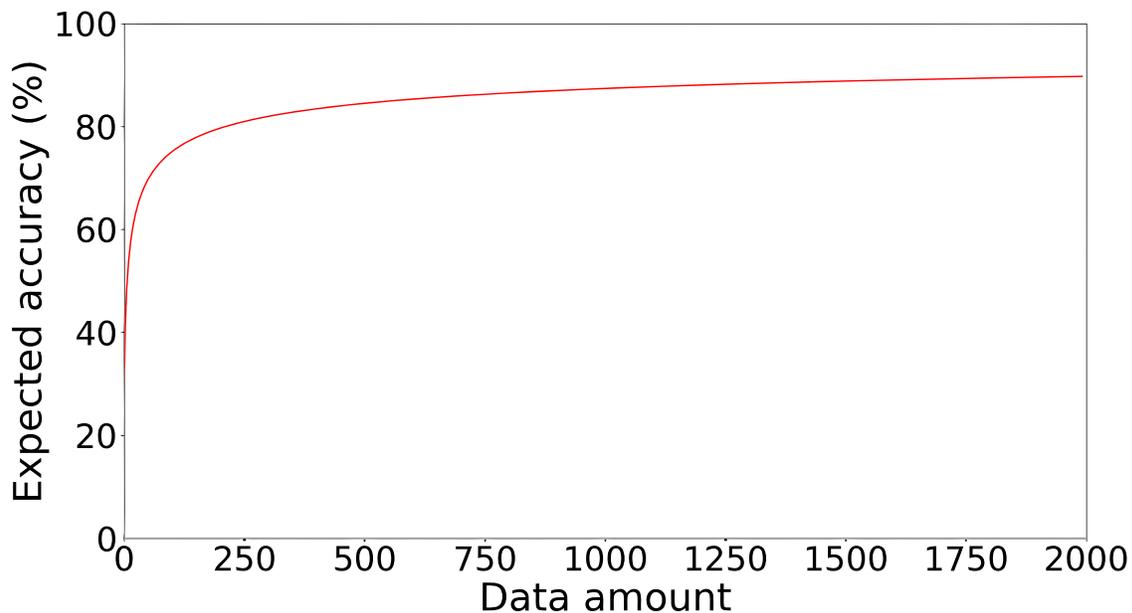}
    \caption{Expected accuracy.}
	\label{fig5}
\end{figure}

In this section, we have demonstrated the performance of the scheme through extensive simulation experiments. The simulations are trained based on the publicly available MNIST dataset. We assume that there are nine edge servers in the network, each of which is equipped with a cache queue, and our simulation results are obtained by averaging the results within the nine edge servers. The initial time slot $t_0$ is 0. The energy of vehicle $k$ in the initial time slot $t_0$ is a random value between 50 and 100. The duration of each time slot is $50ms$. We assume that there are 100 vehicles in each edge server and each vehicle carries 1000 data. The selected vehicles send 10 data to the edge server in each time slot. Each data are an image sample with 1 $MB$ size. The parameters used in the simulation are listed in Table~\ref{tab2}. The utility function of expected accuracy $U(n(t))$ is modeled by the generic learning curve \cite{50}, which is shown in Figure~\ref{fig5}. The expected accuracy when the amount of training data are $x$ is calculated as

\begin{equation}
E_{acc}(x) = 1-l_{rate} \cdot x^{d_{rate}}, 
\label{eq33}
\end{equation}
where $l_{rate}$ is the learning rate and $d_{rate}$ is the decay rate.

\begin{table}[!htbp]
\caption{Related parameters values.}
\label{tab2}
\centering
\setlength{\tabcolsep}{2mm}
\begin{tabular}{ccccccccc} 
\hline\hline\noalign{\smallskip}
\textbf{Parameter} &\textbf{Value} &\textbf{Parameter} &\textbf{Value}\\
\hline
$K$ & 100 & $Q_{max}$ & 2000 MB \\
$D$ & 1000 m & $R$ & 500 m \\
$V$ & $10^{10}$ & $h$ & 50 m \\
$f$ & 2.5 GHz & $Noise$ & 0-1 dB \\
$N$ & 3 & $\alpha$ & $5.6\times10^6$ \\
$\delta^2$ & 7.5 dB & $C_n$ & 10 MB \\
$\sigma^2$ & 0.7 & $\mu$ & 15 m/s \\
$v_{\max}$ & 16.4 m/s & $v_{\min}$ &  13.6 m/s\\
$l_{rate}$ & 1 & $d_{rate}$ &  -0.3\\
\hline\hline\noalign{\smallskip}
\end{tabular}
\end{table}

To evaluate our proposed vehicle selection scheme, we compared it with three other vehicle selection schemes \cite{20}, i.e.,

\noindent \textbf{Maximum selection scheme}: The edge server receives data from all vehicles in each time slot \cite{51, 52}.

\noindent \textbf{Static selection scheme}: The edge server randomly selects the same number of vehicles in each time slot. In the simulation, 5 vehicles are selected randomly to transmit data in each time slot \cite{53}.

\noindent \textbf{Random selection scheme}: The number of selected vehicles is estimated in the same way as our proposed scheme; however, it selects vehicles randomly without considering different vehicles' resource statuses \cite{52, 53, 54}.

\begin{figure}[htbp]
    \centering
    \includegraphics[width=1\textwidth]{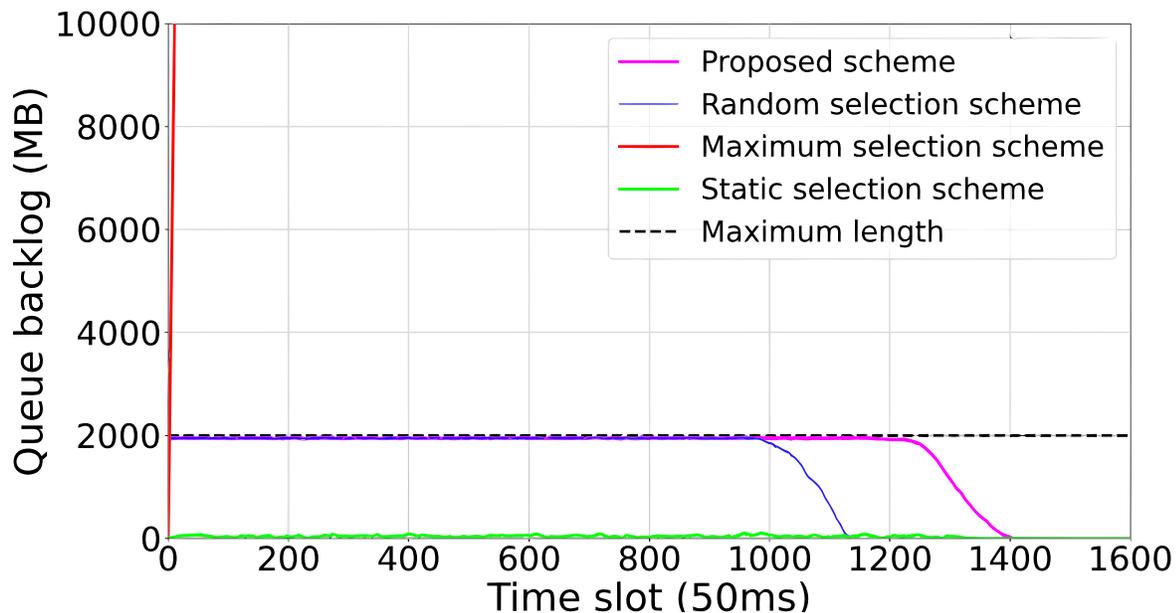}
    \caption{Queue backlog.}
	\label{fig6}
\end{figure}

Figure~\ref{fig6} compares the queue backlog under different selection schemes. The black dashed line in the figure is the maximum queue length $Q_{max}$. It can be seen that the queue backlog rises rapidly and exceeds $Q_{max}$ when the maximum selection scheme is adopted, which indicates that the queue of the edge server is overflowed and thus becomes unstable. This is because that the maximum selection scheme selects all the vehicles to upload data in each time slot. When the static selection scheme is adopted, the queue backlog is much smaller than $Q_{max}$. Although the stability of the queue can be ensured, there are few uploaded data for training, incurring a very low usage of the cache queue. This is because that the static selection scheme selects a small number of vehicles to upload data in each time slot. The queue backlog under both our proposed scheme and the random selection scheme almost keeps $Q_{max}$ before time slot 1000. After that the queue backlog under the random selection scheme gradually decreases, but the queue backlog under our proposed selection scheme still keeps the $Q_{max}$ and then decreases after time slot 1250, hence our proposed scheme ensures the stability for a longer time as compared to the random selection, which validates the performance of our scheme. This is because that both the random selection scheme and our proposed scheme employ the same method to choose the optimal number of vehicles to enable the backlog of the cache queue keeps $Q_{max}$. However, the random selection scheme randomly selects vehicles without considering their resource statuses, which results in some vehicles become unavailable more early than our scheme due to the lack of remaining energy and survivability, thus the edge server cannot collect enough data to maintain the backlog of the queue at $Q_{max}$.

\begin{figure}[htbp]
    \centering
    \includegraphics[width=1\textwidth]{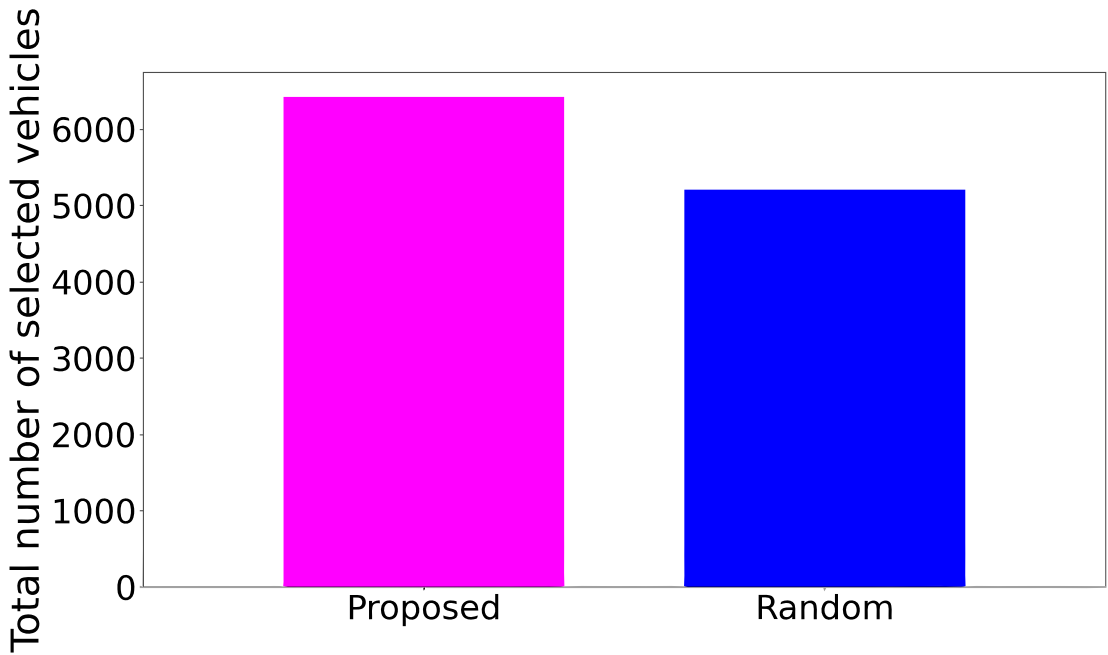}
    \caption{The total number of selected vehicles.}
	\label{fig7}
\end{figure}

Figure~\ref{fig7} compares the total number of selected vehicles under different selection schemes. The total number of selected vehicles in our proposed scheme is 6426, while the total number of selected vehicles in the random selection scheme is 5210. This is because that our scheme can select more vehicles than the random selection scheme during the training due to the consideration of vehicles' resource statuses.

\begin{figure}[htbp]
    \centering
    \includegraphics[width=1\textwidth]{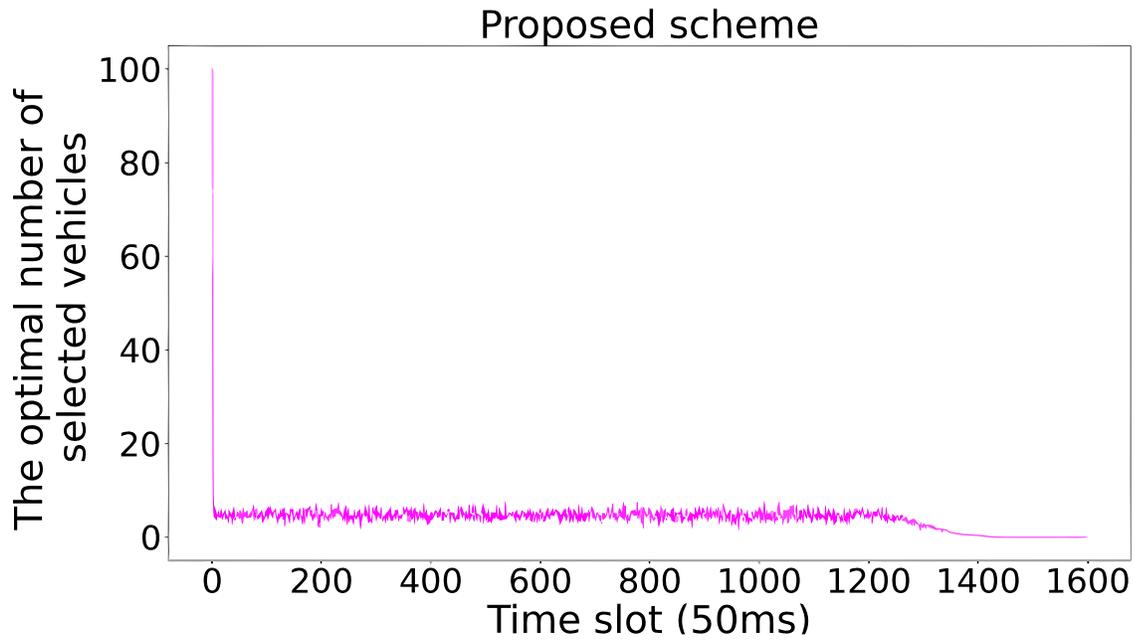}
    \caption{The optimal number of selected vehicles for our proposed scheme.}
	\label{fig8}
\end{figure}

\begin{figure}[htbp]
    \centering
    \includegraphics[width=1\textwidth]{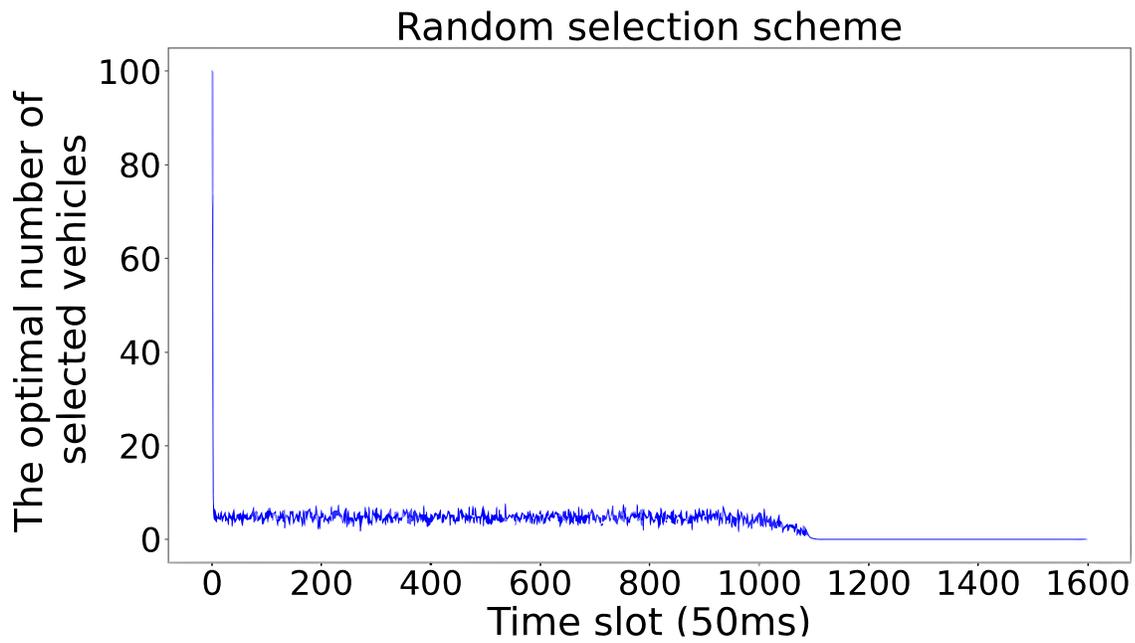}
    \caption{The optimal number of selected vehicles for random selection scheme.}
	\label{fig9}
\end{figure}

Figures~\ref{fig8} and~\ref{fig9} compare the optimal number of selected vehicles in each time slot under different selection schemes. It can be seen that our proposed scheme and the random selection scheme select a similar optimal number of selected vehicles before time slot 1100. However, the optimal number of selected under the random selection scheme is decreased to 0 when the time slot is 1100, while the optimal number of selected under the proposed scheme is not decreased to 0 until time slot 1350. This is because that the random selection scheme does not consider different vehicles' resource statuses and randomly selects vehicles, but our proposed scheme selects vehicles based on their resource statuses, which prolongs the time that the available vehicles exist in the network, thus it can select the optimal number of vehicles for a longer time duration.

It can be seen that the training accuracy with our proposed scheme and that with the random selection scheme basically overlap before time slot 1137. This is because that both of them estimate the optimal number of selected vehicles in each time slot through Lyapunov control theorem. However, the training for the random selection scheme stops at time slot 1137, but for our proposed scheme, the training would not stop until the time slot 1400. This is because that our proposed scheme considers the vehicles' resource statuses to select more vehicles to upload data, which causes more data are involved in the training. In addition, the training accuracy with our proposed scheme is higher than that with the static selection scheme. Moreover, the convergence of the three schemes is essentially the same, because all the three schemes converge when the training contains a sufficient amount of data. The results above show that, our proposed scheme is better than the random selection scheme and the static selection scheme.

\begin{figure}[htbp]
    \centering
    \includegraphics[width=1\textwidth]{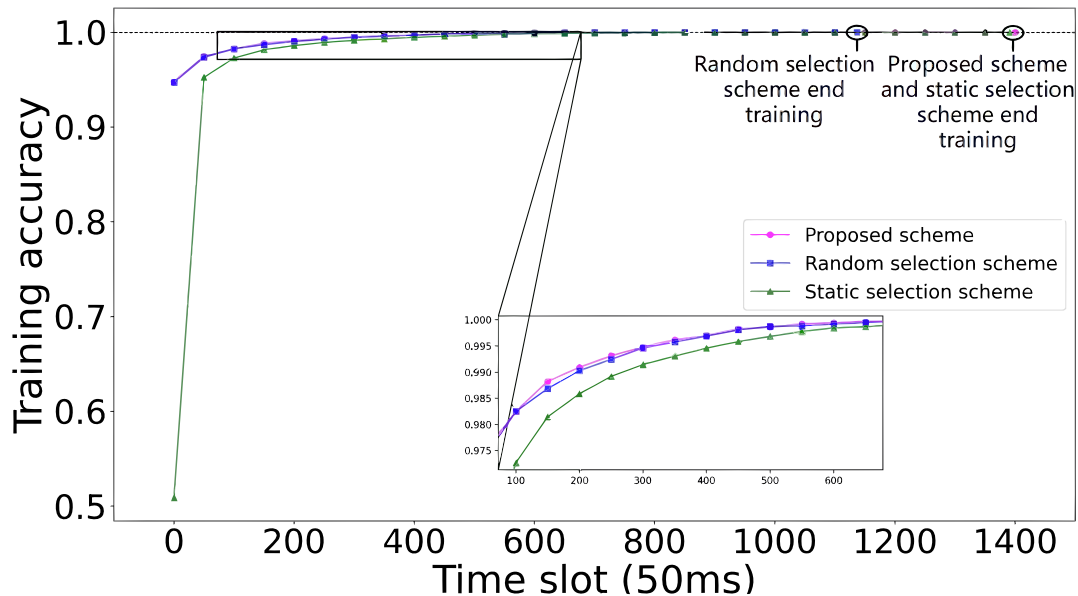}
    \caption{Training accuracy.}
	\label{fig10}
\end{figure}

\begin{figure}[htbp]
    \centering
    \includegraphics[width=1\textwidth]{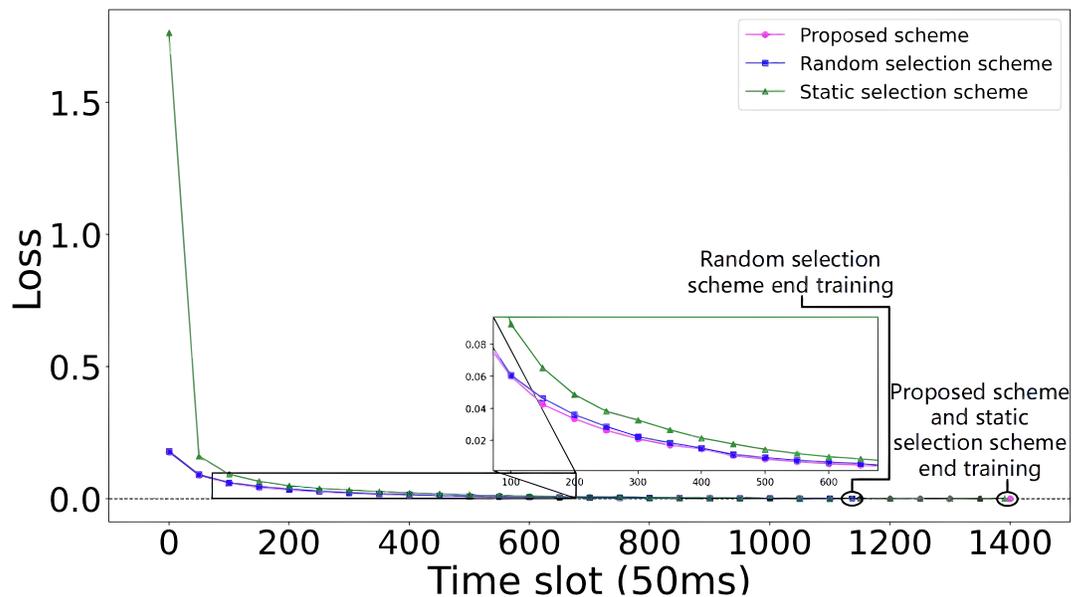}
    \caption{Training loss.}
	\label{fig11}
\end{figure}

The training loss under our proposed scheme and the random selection scheme basically overlap before time slot 1137. This is because both of them estimate the optimal number of selected vehicles in each time slot through Lyapunov control theorem. Later, the training loss under our proposed scheme is lower than that under the static selection scheme.

\section{CONCLUSION}
\label{CONCLUSION}
In this paper, we considered a federated edge learning system for vehicular networks and proposed a new vehicle selection scheme for each edge server, which not only ensures the stability of the edge server's cache queue but also improves the learning model accuracy where more vehicular statuses, including remaining data amount, communication quality, remaining energy and survivability are taken into account. Simulation results demonstrated that our proposed scheme can significantly improve learning accuracy while ensuring the queue stability as compared to other baseline schemes.

\bibliographystyle{IEEEtran}
\bibliography{myref}

\end{document}